# Photonic Interference Cancellation with Hybrid Free Space Optical Communication and MIMO Receiver

Taichu Shi, Yang Qi, and Ben Wu*
*Department of Electrical and Computer Engineering, Rowan University, Glassboro, NJ 08028, USA*

**Abstract:** We proposed and demonstrated a hybrid blind source separation system which can switch between multiple-input and multi-output mode and free space optical communication mode depends on different situation to get best condition for separation.

## 1. Introduction

Blind source separation (BSS) is an effective way to separate the signal of interest (SOI) without knowing the original source signals. It can reduce the interference in wireless communication, especially in the current development of wireless communication technology. Traditional BSS usually uses a multiple-input and multiple-output (MIMO) receiver to implement. It treats the mixed signals received by different receivers as linear combinations from different signal sources. This combination is usually called a mixing matrix, and the signals from receiver can be regarded as the product of the transmitting source signals and the mixing matrix. Therefore, by observing and calculating the signals from the receiver, we can find the de-mixing matrix and then separate the SOI from interference. Condition number is a property of mixing matrix which represents the separability of mixed signals. When the receivers are physically closer to each other, the condition number will increase making it more difficult to separate. This situation is very common in mobile terminals such as mobile phones.

Free space optical communication (FSO) is a line-of-sight wideband communication technology. Its general principle is to modulate and transmitter the laser, and receive the signal at the receiver and then convert it into an electrical signal. If we use FSO to transmit and receive the reference for cancellation, the condition number can be greatly reduced. Since FSO is a line-of-sight wideband communication technology, the contradiction between transmission distance and signal quality is very prominent. When the transmission exceeds a certain distance, the beam will widen, making it difficult for the receiving point to receive correctly.

In this paper, we proposed and demonstrated a hybrid blind source separation system which can switch between MIMO mode and FSO mode depends on different conditions. By building a hybrid FSO and MIMO system, it can complement the advantages of two modes. It can solve the problem that the de-mixing matrix cannot be found correctly because the receiving antennas are physically close to each other.

## 2. Principle and Setup

Fig. 1 shows the schematic diagram of BSS system. The BSS is implemented by a MIMO receiver. The mixed signals are represented by [1, 2]:

$$\boldsymbol{X} = \boldsymbol{A}\boldsymbol{S} \text{ or, } \begin{bmatrix} x_1 \\ x_2 \end{bmatrix} = \begin{bmatrix} a_{11} & a_{12} \\ a_{21} & a_{22} \end{bmatrix} \begin{bmatrix} s_{soi} \\ s_{int} \end{bmatrix} \tag{1}$$

In this formula, $\boldsymbol{A}$ represents the mixing matrix. $x_1$ and $x_2$ are received mixed signals from MIMO antennas. The mixing matrix describes how SOI and interference are mixed. To separate the SOI from interference, we need to find the de-mixing matrix $\boldsymbol{A}^{-1}$. The difficulty of separation is positively correlated to the condition number of the matrix $\boldsymbol{A}$ [3]:

$$cond\ (\boldsymbol{A}) \equiv \|\boldsymbol{A}\| \cdot \|\boldsymbol{A}^{-1}\| \tag{2}$$

$$\|\boldsymbol{A}\| \equiv \max_j \sum_{i=1}^n |a_{ij}| \tag{3}$$

If the $cond\ (\boldsymbol{A})$ has a very large value, it is called ill-conditions [3]. It often exists in mobile terminals where the two receiving antennas are physically close to each other. From Eq. 1, we can see that $x_1 = a_{11}s_{soi} + a_{12}s_{int}$ and $x_2 = a_{21}s_{soi} + a_{22}s_{int}$. If the two antennas are very close, in that case, the two receiving antennas will receive similar signals where $a_{11} \approx a_{21}$ and $a_{12} \approx a_{22}$, which will make blind source separation very difficult to perform. In the extreme condition, if the $a_{11} = a_{21}$ and $a_{12} = a_{22}$, the two antennas will have same mixed signal containing SOI and interference that we can not separate and the condition number $\boldsymbol{A}$ will become infinite.

When we adding a FSO channel into the system, one of the MIMO antennas is replaced by the FSO link. In this experiment, we send the reference signals generated by an arbitrary waveform generator used as an interference generator. The FSO reference includes interference for cancellation which makes the $a_{21}$ in Eq.1 become 0. Thus, the new mixing matrix can be written in:

$$\begin{bmatrix} x_1 \\ x_2 \end{bmatrix} = \begin{bmatrix} a_{11} & a_{12} \\ 0 & a_{22} \end{bmatrix} \begin{bmatrix} s_{soi} \\ s_{int} \end{bmatrix} \quad (4)$$

Therefore, the $a_{21}$ and $a_{22}$ will have significate difference and solve the ill-condition problem making it easier for blind source separation. Fig. 1 (b) shows the experimental setup for ill-condition. Here we put two transmitting antennas relatively far from each other and two receiving antennas are very close. The distance between two receiving antennas is 2cm. Tx1 represents the SOI, it's 111cm from the receivers. Tx2 represents the interference, it's 60cm from the receivers.

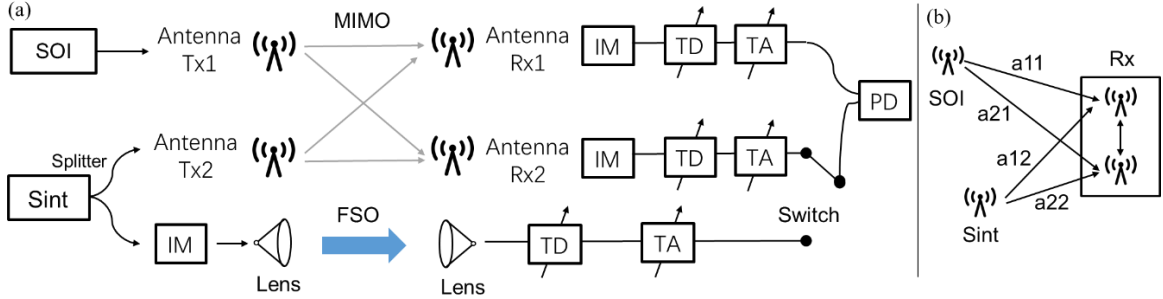

Fig. 1. (a)BSS system schematic (IM: optical intensity modulator, TD: tunable optical delay, TA: tunable optical attenuator, PD: photodetector) (b) Antenna positions.

### 3. Results and Analysis

The results are shown in Fig. 2. Fig. 2 (a) and (c) are the transmission spectrum between 839 and 841MHz (The best transmission frequency band of the antenna) of in MIMO mode. The parameters from same transmitter to different receivers are very similar ($a_{11} \approx a_{21}$ and $a_{12} \approx a_{22}$). Fig. 2 (b) and (d) are the transmission spectrum when switching to FSO mode. We can see that $a_{21}$ drops to 0 and $a_{12}$ has a significant difference from $a_{12}$. Fig. 2 (b) shows the constellation diagram s of recovered SOI between FSO mode and MIMO mode. The SOI we used here are QPSK, 16QAM and 64QAM signals. Clear constellation diagram can be obtained when switching to FSO mode, where in the MIMO mode, if the modulation order increases, we cannot have a clear constellation diagram.

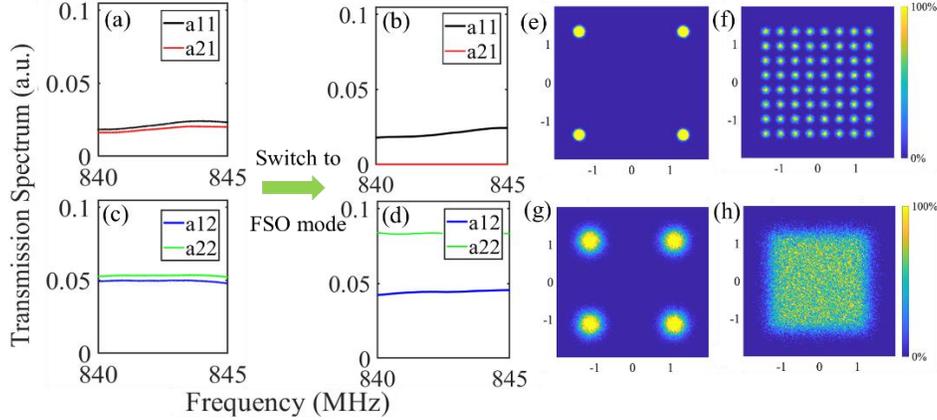

Fig. 2. (a)-(d) Transmission spectrum from transmitters to receivers before and after FSO mode (e)(f) Different constellation diagrams of different modulation in FSO mode (g)(h) Same signals in MIMO mode. The constellation become blurry and unidentifiable in some cases.

### 4. Conclusion

We proposed and experimentally demonstrated a hybrid blind source separation system by building a hybrid FSO and MIMO system. The system then can complement the advantages of FSO and MIMO. FSO mode can greatly reduce the condition number of mixing matrix for easier separation, while the MIMO mode ensure the system is robust and reliable when the transmission path of FSO is blocked.